\documentclass[a4paper,11pt]{article}

\usepackage{jinstpub} 

\usepackage{graphicx}
\usepackage{caption}
\usepackage{subcaption}

\title{Machine learning-based method of calorimeter saturation correction for helium flux analysis with DAMPE experiment}

\author[a,1]{Mikhail Stolpovskiy,\note{Corresponding author.}}
\author[a]{Xin Wu,}
\author[a]{Andrii Tykhonov,}
\author[a]{Maksym Deliyergiyev,}
\author[a]{Chiara Perrina,}
\author[a]{Maria Munoz,}
\author[a]{David Droz,}
\author[a]{Arshia Ruina,}
\author[b,c]{Enrico Catanzani,}

\affiliation[a]{University of Geneva, CH-1205 Geneva, Switzerland}
\affiliation[b]{Istituto Nazionale di Fisica Nucleare (INFN) - Sezione di Perugia, I-06123 Perugia, Italy} \affiliation[c]{Dipartimento di Fisica e Geologia, Universita‘ di Perugia, I-06123, Perugia, Italy}

\emailAdd{mikhail.stolpovskiy@unige.ch}

\collaboration[c]{on behalf of the DAMPE collaboration}

\abstract{DAMPE is a space-borne experiment for the measurement of the cosmic-ray fluxes at energies up to around 100 TeV per nucleon. At energies above several tens of TeV, the electronics of DAMPE calorimeter would saturate, leaving certain bars with no energy recorded. In the present work we discuss the application of machine learning techniques for the treatment of DAMPE data, to compensate the calorimeter energy lost by saturation.}


\begin{document}
\maketitle
\flushbottom

\section{Introduction}
Cosmic rays (CR), in a hundred years after their discovery, still keep lots of mysteries, including the questions about their origin and propagation \cite{cr_origin, cr_propa}. To address these questions, the fluxes of different constituents of the CRs are measured with high precision by spectrometer instruments up to GV rigidities \cite{cr_measurements}. However, yet more precision at highest energies is required to reduce the statistical and systematic uncertainties in those measurements.

DAMPE \cite{dampe1, dampe2} is one of the leading instruments in direct measurements of the CR, electron/positron and $\gamma$-ray fluxes. It was launched on December 17, 2015 onto a Sun-synchronous orbit at an altitude of 500 km. DAMPE has been working smoothly since then for more than 6 years. The DAMPE detector consists of (in order of detection of a CR):

\begin{itemize}
    \item Plastic Scintillator strip Detector (PSD) \cite{psd1, psd2},
    \item Silicon-Tungsten tracKer–converter (STK) \cite{stk1, stk2},
    \item BGO imaging calorimeter (BGO) \cite{bgo1, bgo2},
    \item NeUtron Detector (NUD) \cite{nud}.
\end{itemize}

Among DAMPE's recent results are the measurements of proton \cite{proton}, helium \cite{helium} and electron/positron \cite{electron} fluxes. Preliminary results on $\gamma$-ray observations were also presented \cite{gamma}. These studies show interesting spectral features which help to push our understanding of the CR astrophysics and constrain dark matter models \cite{physics1, physics2, physics3}.

The main sub-detector of DAMPE, the BGO calorimeter, consists of 14 layers of Bismuth-Germanium-Oxide crystal bars, with 22 bars (25 $\times$ 25 $\times$ 600 mm$^3$) in each layer \cite{bgo1}. Each consecutive layer is oriented orthogonally to the previous one, giving a possibility to measure the CR showers in three dimensions. In total, it is 31.5 radiation length thick, thus it is able to fully contain an electromagnetic shower. It has a nuclear interaction length thickness of 1.6. The fluorescence light is read by two photo-multipliers, glued up on both ends of each bar.

The readout electronics of BGO bars is not capable of measuring energy depositions of more than several TeV. In case of a higher energy deposition, the PMT electronics saturate and record zero signal. This effect reduces the recorded energy of the shower. It weakens the connection between the true kinetic energy of a CR and its recorded energy, which, at the end of the day, increases the systematic uncertainties of the reconstructed flux. An accurate reconstruction of the energy missed due to saturation is required to mitigate this effect. Later in the paper we call such reconstruction a saturation correction.


\section{BGO calorimeter} \label{BGO_calorimeter}

The scintillation light created by the CR shower in the BGO bars is detected by two PMTs coupled to the two ends of the bar. This coupling is made using two different optical filters S0 and S1. The filters are adjusted in such a way that the vertical muon MIP gives signal, shown in table \ref{tab:Yifeng}. The table gives only approximate values for the MIP signal since it actually depends on many parameters (for example temperature).

\begin{table}[]
\centering
\begin{tabular}{|l|ll|}
\hline
Layer & \multicolumn{1}{l|}{S0} & S1            \\ \hline
1     & $\sim$500 ADC           & $\sim$500 ADC \\ \cline{1-1}
2     & $\sim$500 ADC           & $\sim$200 ADC \\ \cline{1-1}
3-12  & $\sim$500 ADC           & $\sim$100 ADC \\ \cline{1-1}
13    & $\sim$500 ADC           & $\sim$200 ADC \\ \cline{1-1}
14    & $\sim$500 ADC           & $\sim$500 ADC \\ \hline
\end{tabular}
\caption{Attenuation settings of the optical filters on the S0 and S1 PMTs, glued to both ends of each BGO bar. The values given in the table correspond to the approximate signal from a vertical muon MIP.}
\label{tab:Yifeng}
\end{table}

One can see that in the center of the BGO calorimeter the optical filters on the S1 end of each bar are less transparent, so the signal in S1 PMT is about 5 times weaker than it is on the other end. Thus the S1 end PMT is capable of recording higher energy deposits. For the average energy deposit in a bar, we have two independent measurements from S0 and S1 sides, which increases the energy resolution of BGO calorimeter in the whole. The upper and lower layers have different settings in order to enhance the energy and spatial resolution at the beginning and at the end of a shower. This feature is particularly important for the electron discrimination, which relies on the shower imaging precisely at the extremities of the shower.

\begin{figure}
    \centering
    \includegraphics[width=0.7\textwidth]{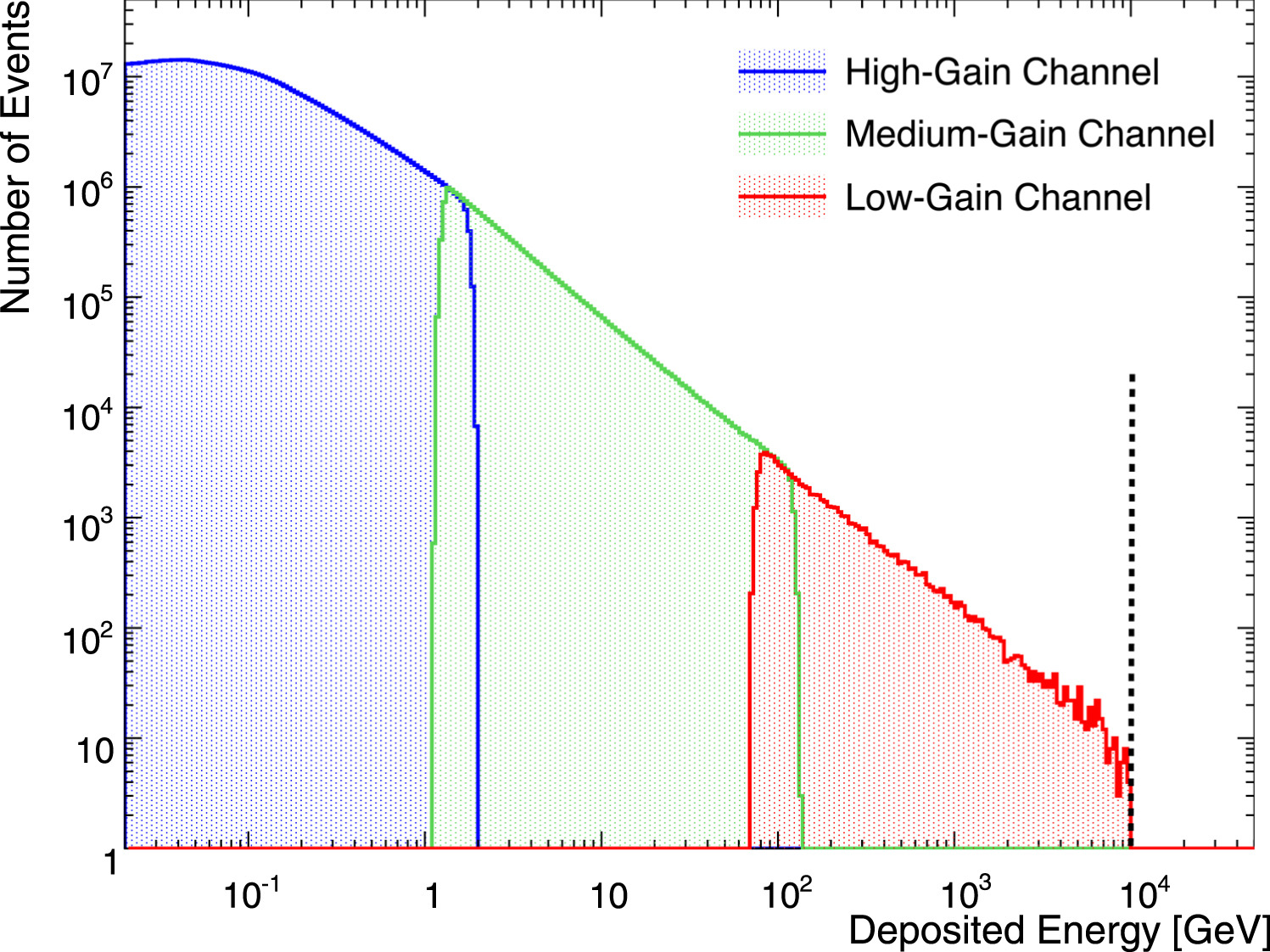}
    \caption{An example of the energy deposit spectrum, reconstructed from the S1 end of one of the BGO bars. Different color histogram correspond to the high-gain channel (blue), medium-gain (green) and low-gain (red). The vertical black line corresponds to the upper limit of the measurement for this bar.}
    \label{fig:gain}
\end{figure}

\begin{figure}
    \centering
    \includegraphics[width=\textwidth]{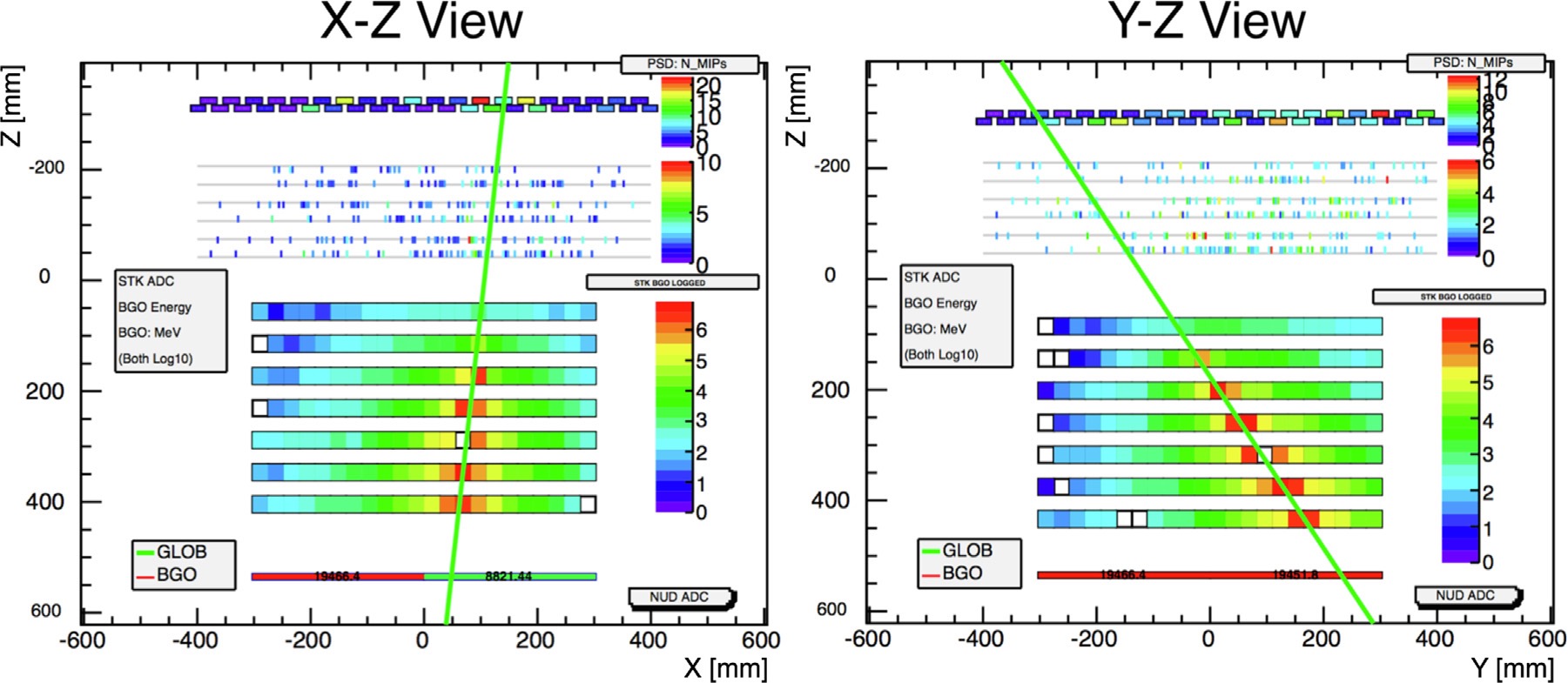}
    \caption{An example of saturated event. Views from two orthogonal directions are shown, corresponding to the coordinate system of DAMPE. The reconstructed direction of the incoming CR is shown with green lines. The sub-detectors are visible: on top the PSD detector, then STK tracker, BGO and NUD on bottom. Color shows the strength of the signal in each channel. The saturated bars are situated right on the shower axis and are shown as zero (white) bars. The off-axis zero bars are not saturated, the signal in them is below the noise level.}
    \label{fig:monitor}
\end{figure}

Each PMT is read out by three sensitive dynodes: 2, 5 and 8, which correspond to the low, medium and high gain channels \cite{ecal}. The high linearity and time stability of the BGO readout system is assured by the on-orbit calibration \cite{bgo_calibration}, see figure \ref{fig:gain}. The upper limit on signal is set for each dynode. The signals recorded beyond this limit are discarded on orbit. When the energy deposit in a bar is so large that it exceeds the limit of the low-gain dynode on the S1 end of the bar, the energy is not recorded and the event becomes saturated. An example of a saturated event with two saturated bars right on the shower axis is shown on the figure \ref{fig:monitor}.

\section{Method}

\subsection{Data selection}

We use Monte-Carlo (MC) data of helium-4 with primary energies from 10 TeV to 500 TeV. We select events where the true primary particle trajectory passes through the whole detector: it passes through the first layer of PSD on one side and through the last layer of BGO, excluding the width of one BGO bar on each of the four sides. Secondly, the event should satisfy the standard event selections used in DAMPE data analysis: the high energy trigger should be fired and the special set of selections designed to exclude the side-in cosmic rays is applied. However, once we apply the selection on the true particle trajectory, the latter selections have nearly 100\% efficiency. The full number of events used for training and evaluation of the machine learning model is over 110 thousands.

The saturated bars can be detected as bars with zero energy registered, adjacent to a bar with registered energy more than 25 GeV.

\subsection{Model architecture}

To reconstruct the energy missed due saturation in the BGO bars we apply two distinct convolutional neural network models (CNN) \cite{CNN}. The first model is dedicated to the reconstruction of the saturation in the last layer of the detector and the second model is for the saturation in the middle layers (with an exception which we will explain later). In the text below we will call these models as last-layer and middle-layer models correspondingly.

\begin{figure}
    \centering
    \begin{minipage}[t]{0.45\textwidth}
        \centering
        \includegraphics[width=\textwidth]{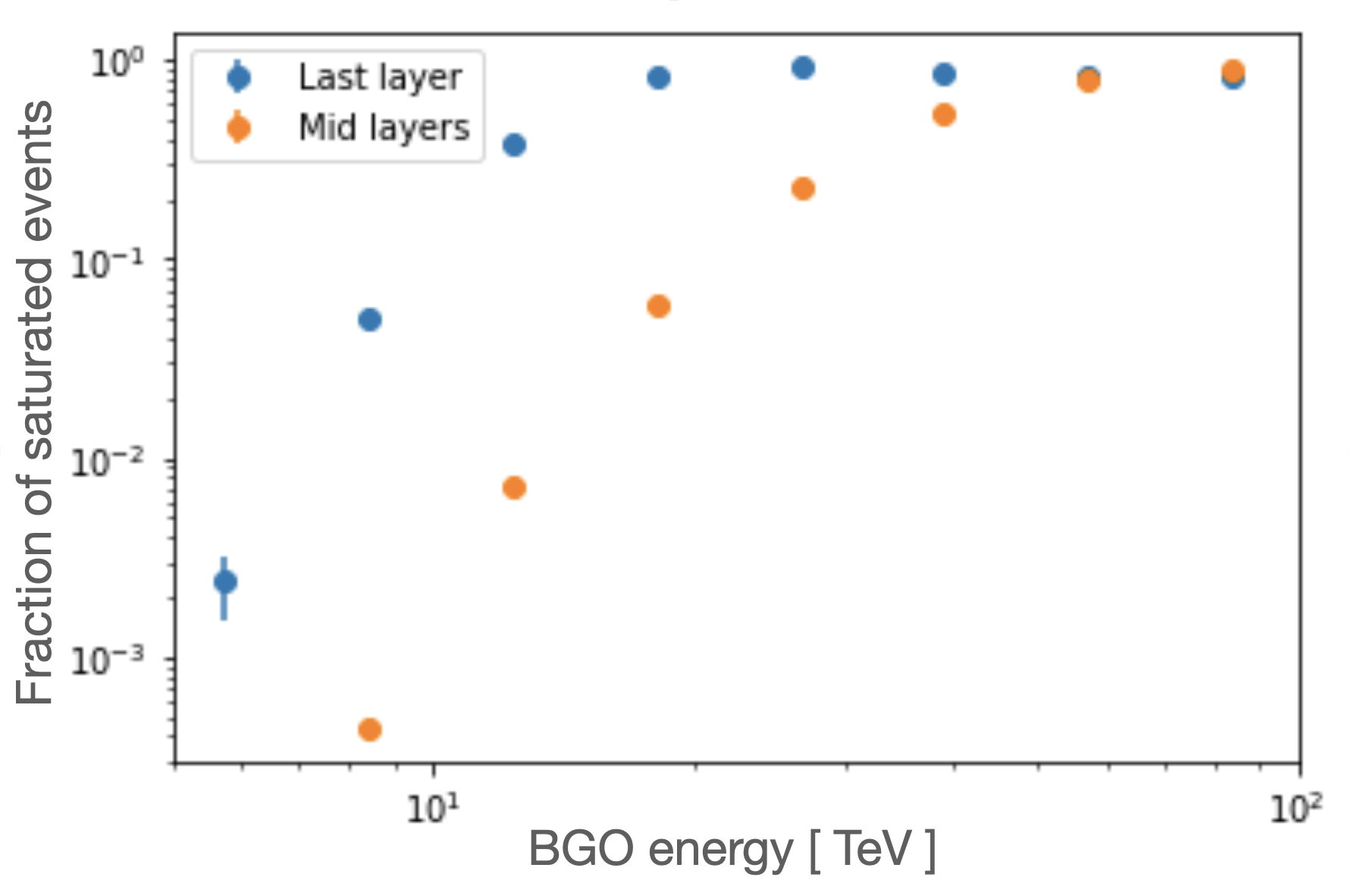}
        \caption{Fraction of events with saturation in the last layer and in the middle layers. Note that since there are events where saturation happens in the last layer and in the middle layers simultaneously, these two classes of events are partially overlapping.}
        \label{fig:frac_sat}
    \end{minipage}
    \hfill
    \begin{minipage}[t]{0.45\textwidth}
        \centering
        \includegraphics[width=\textwidth]{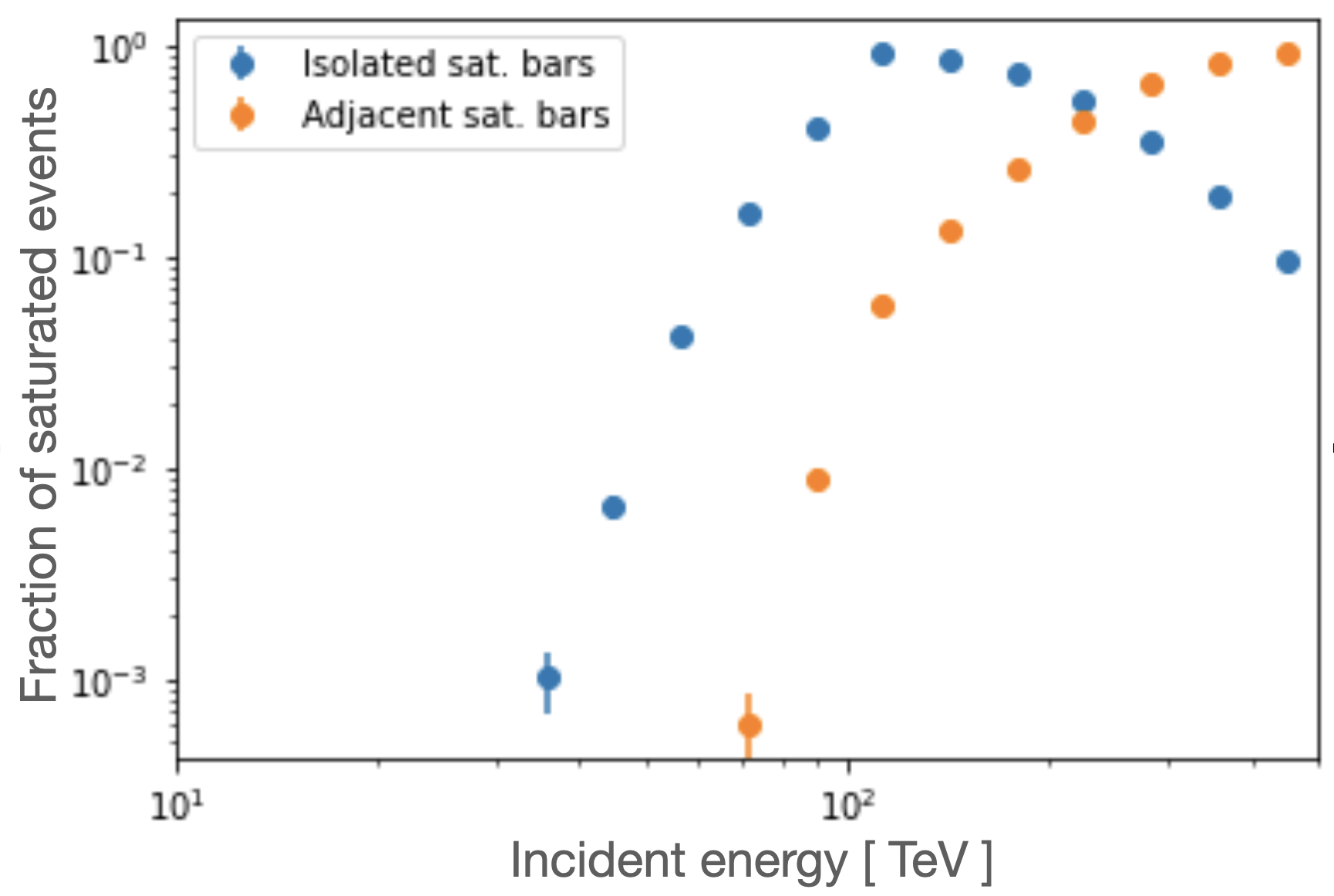}
        \caption{Fraction of events with saturation, for the events exclusively with isolated saturated bars (blue points) and for the events where the adjacent saturated bars are present (orange points), as function of the incident energy of the particle.}
        \label{fig:frac_sat_adj_iso}
    \end{minipage}
\end{figure}

The reason to develop a separate model for the last layer is the different gain settings in the last layer, as mentioned in the section \ref{BGO_calorimeter}. The consequences of this difference are double-fold. First, higher gain leads to a large fraction of events where the saturation happens in one of the last layer bars. Fractions of events with saturation in the last layer and in the middle layers are shown on the figure \ref{fig:frac_sat}. Thus the events with saturation in the middle of BGO are under-represented, especially at energies below one hundred TeV that constitute the bulk of the events, registered with DAMPE. Secondly, the missing energy in the events with saturation in the last layer is substantially lower than the energy lost due saturation in the middle layers. In these conditions we find the approach of splitting the last-layer and middle-layer saturations in two different models to be the most accurate.

At very high energies about 100 TeV and more, the  energy deposited is so large that for a large fraction of events many adjacent bars in the shower core become saturated. The fraction of the saturated events with adjacent and isolated bars is shown on the figure \ref{fig:frac_sat_adj_iso}. If a saturated bar in the last layer is adjacent to another saturated bar, either in the same layer or in the layer above, its missing energy is comparable to the missing energy in the middle layers. Keeping this in mind, to train the last-layer model we select events with an isolated saturated bar in the last layer (not adjacent to any other saturated bar neither in the last layer nor in the layer above). Note that in this selection we don't exclude events with saturation in the middle of the BGO. The target value for the last-layer model is the energy lost in the saturated bar in the last layer in units of TeV.

\begin{figure}
    \centering
    \begin{minipage}[t]{0.45\textwidth}
        \centering
        \includegraphics[width=\textwidth]{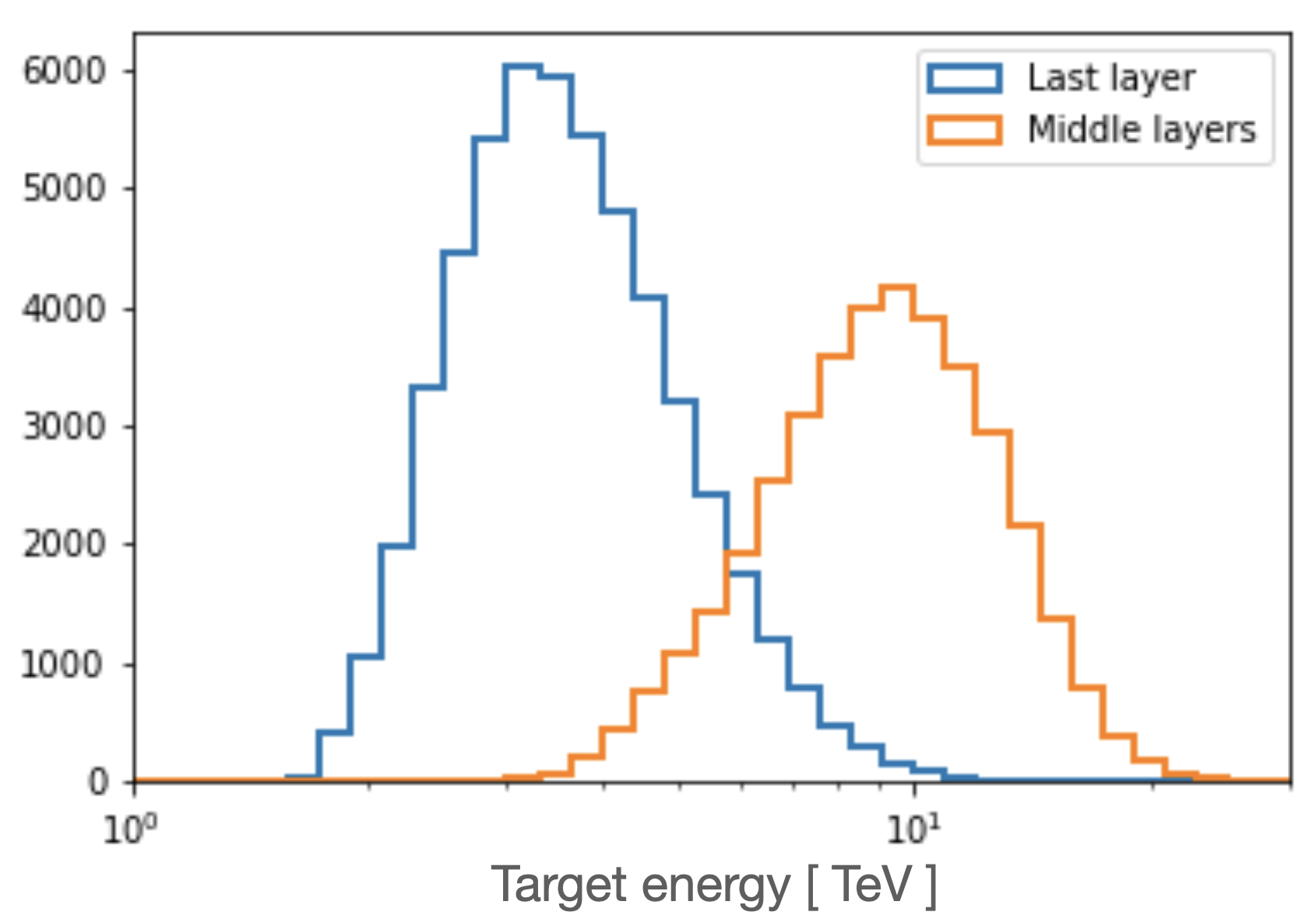}
        \caption{Distribution of the target energy for the last-layer model and the middle-layers model. }
        \label{fig:target_dist}
    \end{minipage}
    \hfill
    \begin{minipage}[t]{0.45\textwidth}
        \centering
        \includegraphics[width=\textwidth]{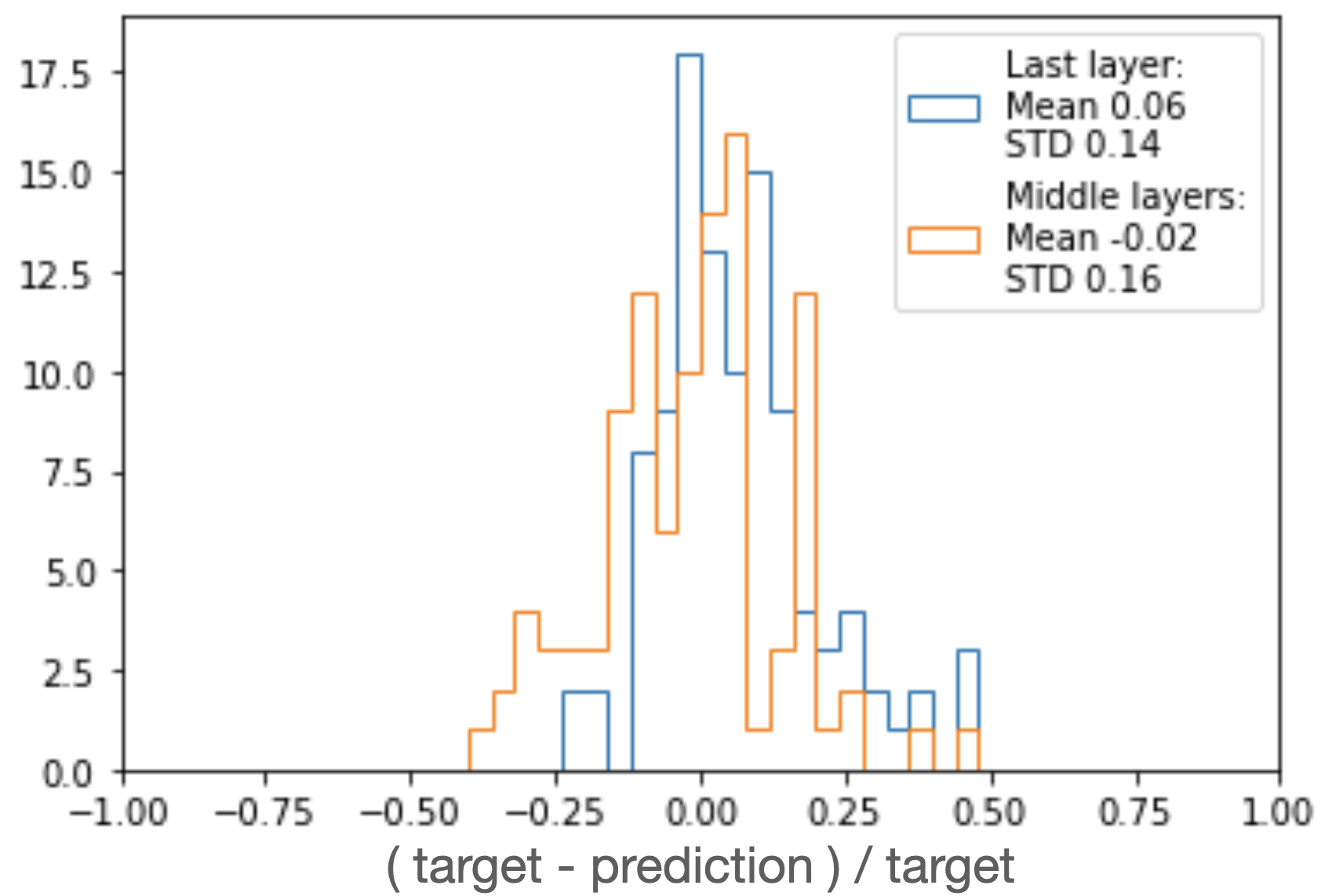}
        \caption{The difference between the reconstructed BGO bar energy deposit and the actual one for the artificially saturated events from the flight data. Blue histogram shows the distribution for the last layer saturated bars, orange one shows it for the middle layer bars.}
        \label{fig:target-prediction_data}
    \end{minipage}
\end{figure}

After the last-layer model is trained and its predictions are obtained, we add these predictions to the correspondent BGO bars. Then the events with the saturation in the middle of BGO, including those with already corrected last layer saturation, are used to train the middle layer model. The target value for the middle-layer model is the energy lost in the saturated bars, divided by the number of saturated bars (that is the average saturated energy per bar) in TeV. Later in the text, the target values for the two models are called target energy. The distributions of target energy for the two classes of events are shown on figure \ref{fig:target_dist}. The number of events used for training and evaluation of the two models is: 53 and 77 thousand for last-layer and middle-layer models, respectively. We split these samples in half, one half is for training of the model and the other half for testing.

\begin{figure}
    \centering
    \includegraphics[width=\textwidth]{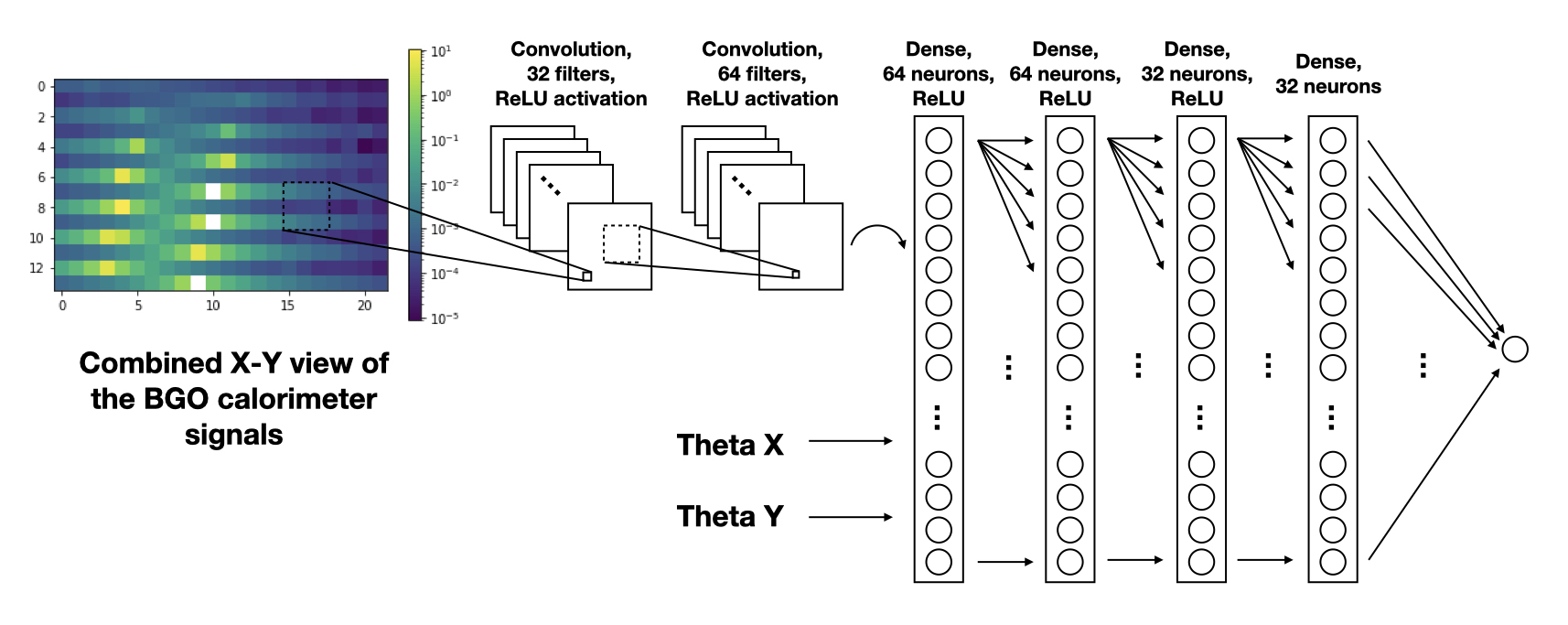}
    \caption{Architecture of the CNN models used in the work. The size of the boxes is proportional to the size of the inputs in each layer. The single dimensional inputs are shown oriented horizontally. The meaning of the colors for different layers is explained in the legend. The last activation layer is for linear activation. The other four use relu activation \cite{relu}.}
    \label{fig:cnn_archi}
\end{figure}

Both models have the same architecture, as shown on figure \ref{fig:cnn_archi}. As inputs we use the combined view of the BGO detector, as shown on the right of figure \ref{fig:cnn_archi}, plus the reconstructed inclination of the arriving cosmic ray in XZ and YZ projections ($\theta_X$ and $\theta_Y$ respectively). The combined view of BGO is constructed of alternating layers from the YZ and XZ sides of the detector. In the example shown in figure \ref{fig:cnn_archi}, one can see three saturated bars, one in the last layer and two in the middle layers, all three isolated from each other. The incident energy of the helium ion is 201 TeV and the inclinations on XZ and YZ projections are -0.3 and -0.2 radian correspondingly. The first part of the neural network consist of two convolution layers, where convolution is applied to the combined view of the BGO. The convolved image is then concatenated with inclination values and fed into a fully connected feed forward neural network \cite{NN} with three hidden layers. The CNN is created, trained and evaluated using TensorFlow \cite{tensorflow} package with a Keras frontend \cite{keras}.

\subsection{Model performance}

\begin{figure}
    \centering
    \begin{minipage}[t]{0.45\textwidth}
        \includegraphics[width=\textwidth]{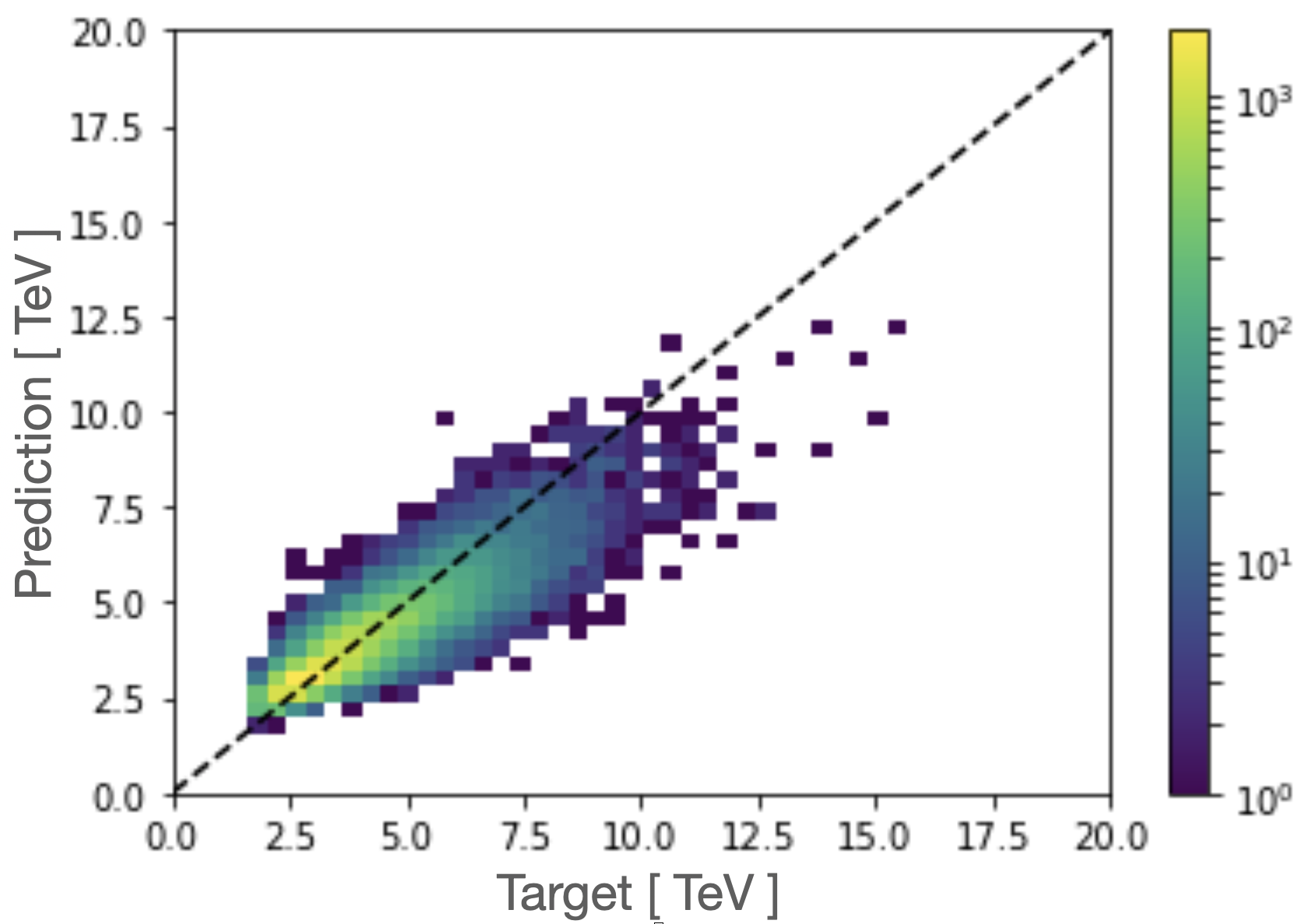}
        \caption{Two-dimentional distribution of reconstructed missing energy (vertical axis) versus true missing energy (horizontal axis) for the saturated bars in the last layer of BGO. Diagonal line of $reconstructed = true$ is shown for reference. Colors correspond to the number of entries in each bin.}
        \label{fig:rec_vs_tar_ll}
    \end{minipage}
    \hfill
    \begin{minipage}[t]{0.45\textwidth}
        \includegraphics[width=\textwidth]{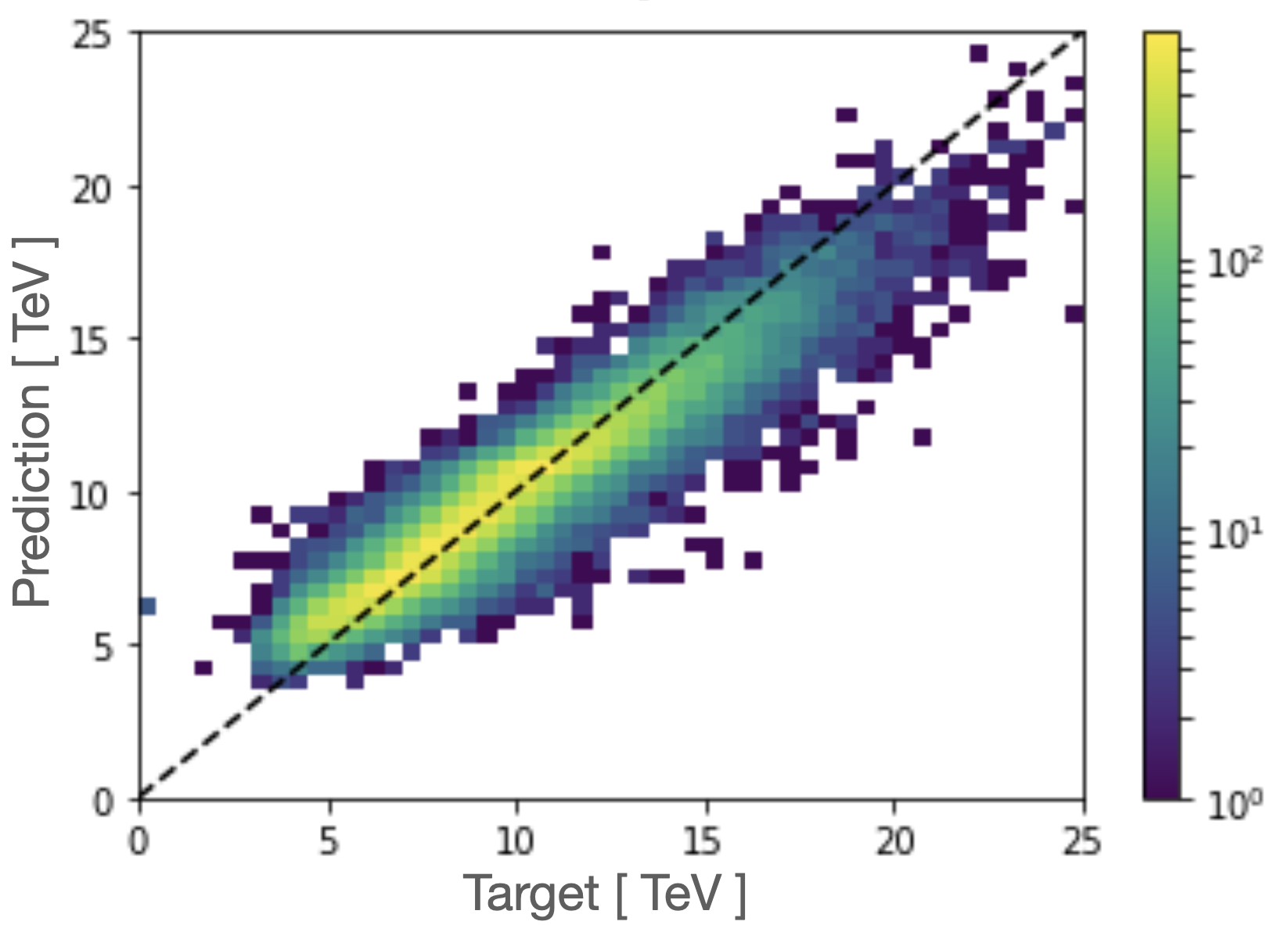}
        \caption{Two-dimentional distribution of reconstructed average missing energy (vertical axis) versus true average missing energy (horizontal axis) for the saturated bars in the middle layers of BGO. Diagonal line of $reconstructed = true$ is shown for reference. Colors correspond to the number of entries in each bin.}
        \label{fig:rec_vs_tar_mid}
    \end{minipage}
\end{figure}

\begin{figure}
    \centering
    \begin{minipage}[t]{0.45\textwidth}
         \includegraphics[width=\textwidth]{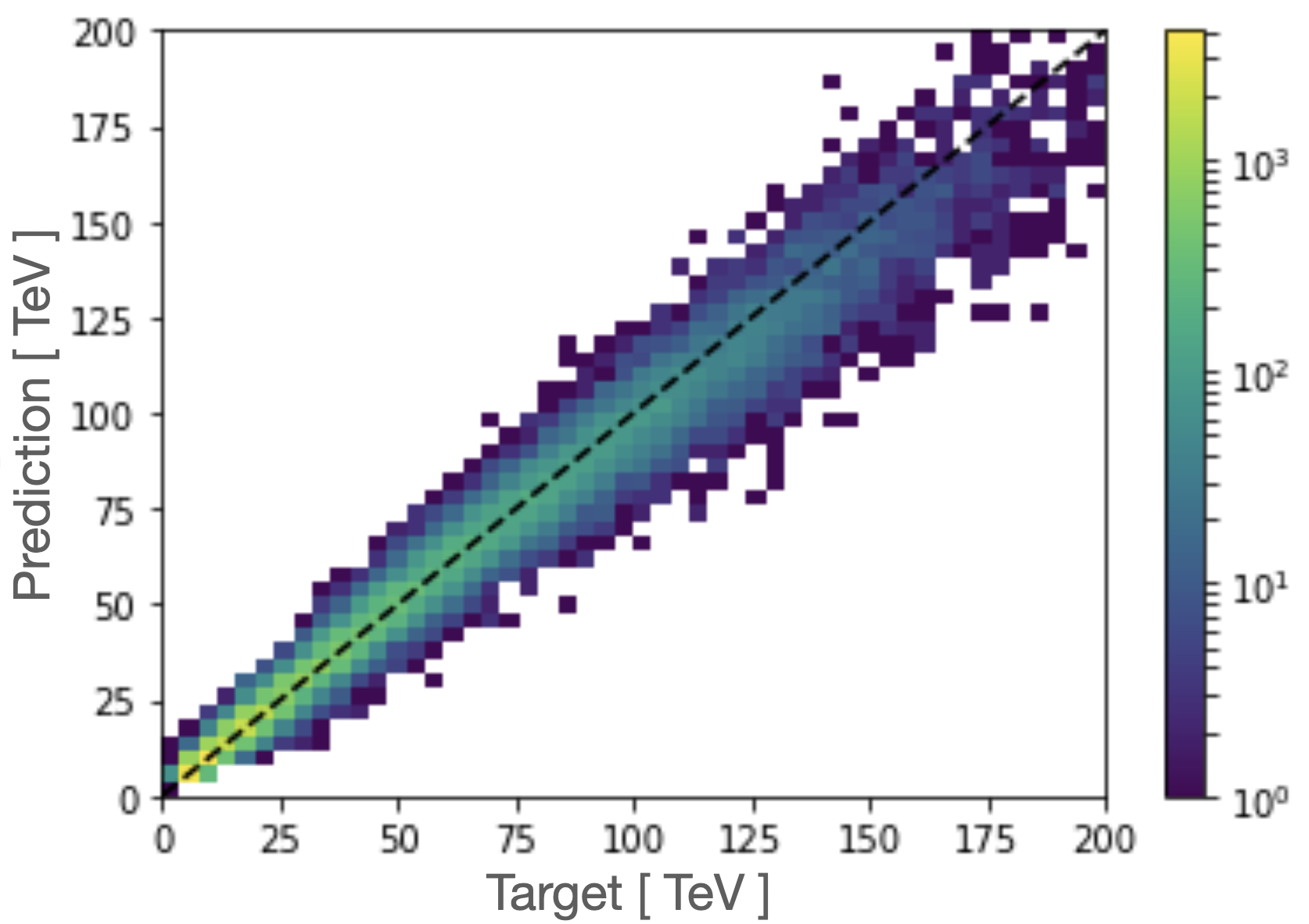}
         \caption{Two-dimentional distribution of reconstructed missing energy versus true missing energy. Diagonal line of $reconstructed = true$ is shown for reference. Colors correspond to the number of entries in each bin.}
         \label{fig:rec_vs_tar_full}
    \end{minipage}
    \hfill
    \begin{minipage}[t]{0.45\textwidth}
        \includegraphics[width=\textwidth]{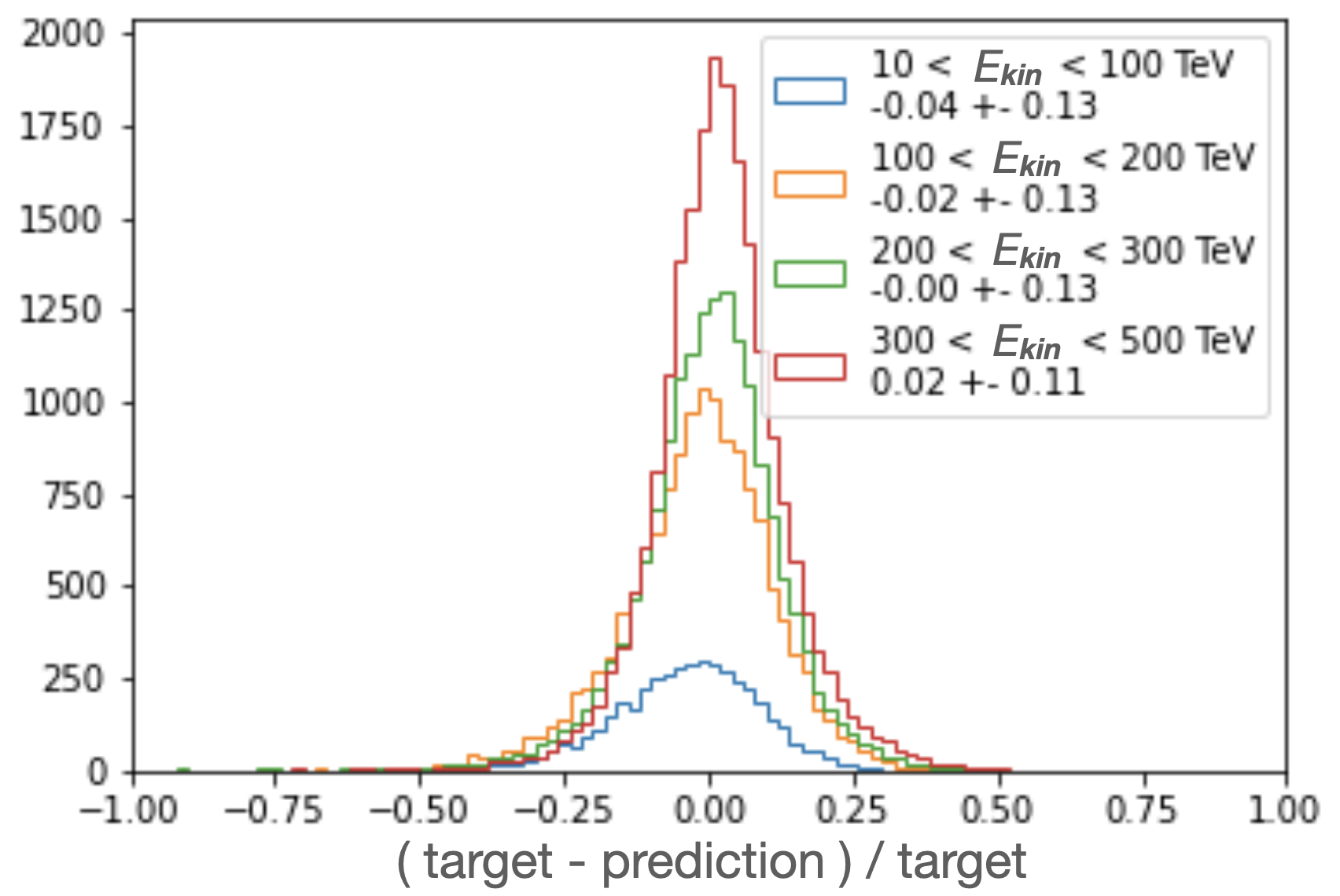}
        \caption{Difference between the true saturation missing energy per event (target) and the reconstructed one (prediction) in different bins of particle primary energy. The mean and standard deviations of the distributions are shown in the legend.}
        \label{fig:target-prediction_primBins}
    \end{minipage}
\end{figure}

The two-dimentional distributions of target energy versus its reconstructed value for the two models are shown in figures \ref{fig:rec_vs_tar_ll} and \ref{fig:rec_vs_tar_mid}. As mentioned above, the full missing energy reconstruction goes in two steps: we reconstruct the missing energy in the last layer, add it to the corresponding bar, and reconstruct the average missing energy in other BGO layers. If the saturated bar in the last layer is adjacent to any other saturated bar, the middle-layer model is directly applied to the full BGO, omitting the first step (same is true for the events without saturation in the last layer). The full reconstructed missing energy versus its true value is shown in figure \ref{fig:rec_vs_tar_full}. One can see that the missing energy in some rare cases reaches 200 TeV, and our model successfully reconstructs it. Apparently, the highest values of missing energy correspond to the largest energies of the primary particle. The distribution of the relative difference between the true saturation missing energy per event (total target) and the reconstructed saturation energy (total prediction) is shown in figure \ref{fig:target-prediction_primBins} in bins of primary energy. The statistical characteristics of the distributions are shown on the plot legend. There is a slight bias towards too high reconstructed energy at primary energies below 100 TeV. This bias can be explained as a well-known effect of bias-variance of the minimisation problems \cite{bias-variance}, it appears here because of the asymmetric distribution of the target energy.

\begin{figure}
    \centering
    \begin{minipage}[t]{0.45\textwidth}
         \includegraphics[width=\textwidth]{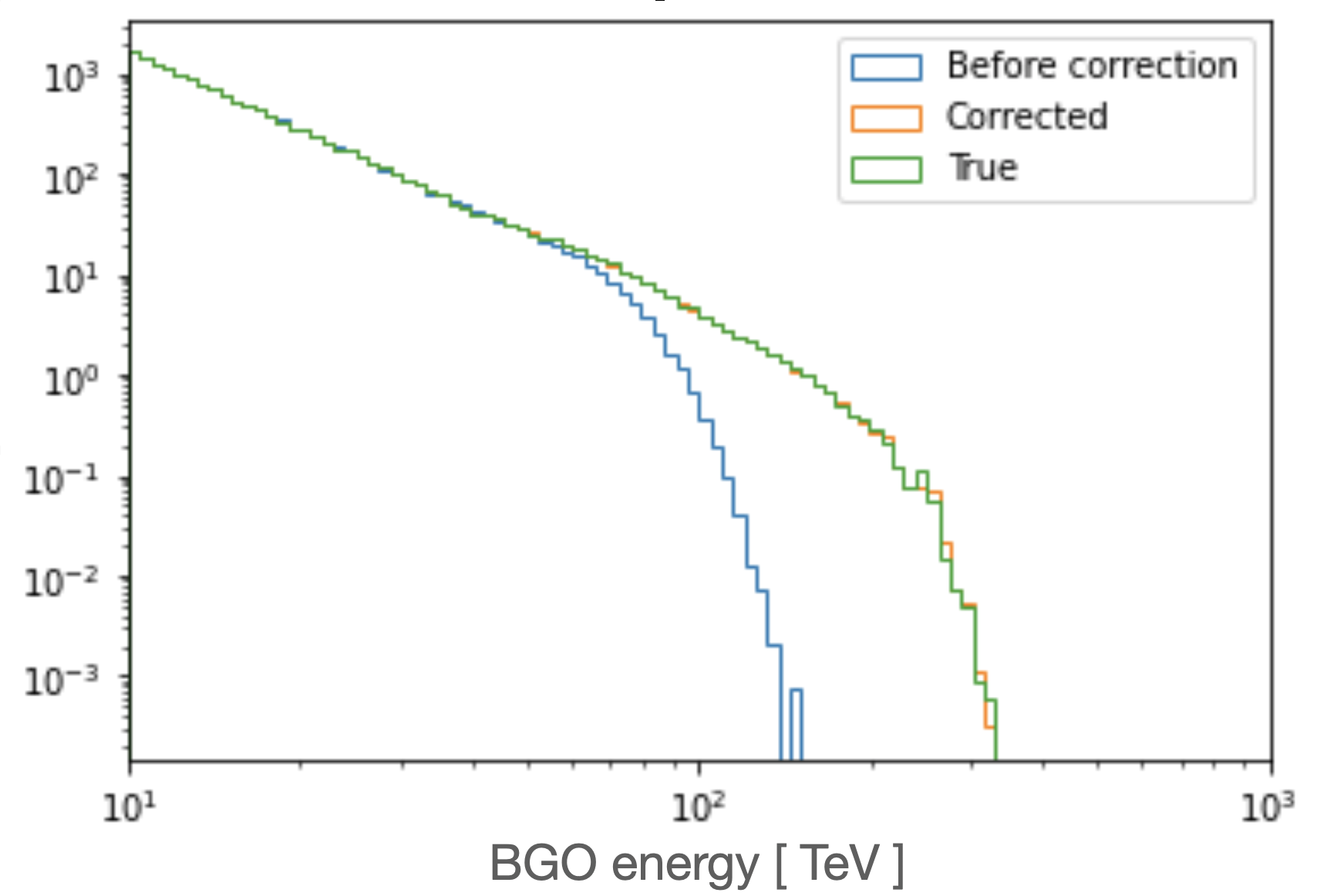}
         \caption{Distributions of total energy in BGO: recorded energy before correction (blue), true deposit energy distribution (green) and recorded energy after correction (orange).}
         \label{fig:BGO_energy_distribution}
    \end{minipage}
    \hfill
    \begin{minipage}[t]{0.45\textwidth}
         \includegraphics[width=\textwidth]{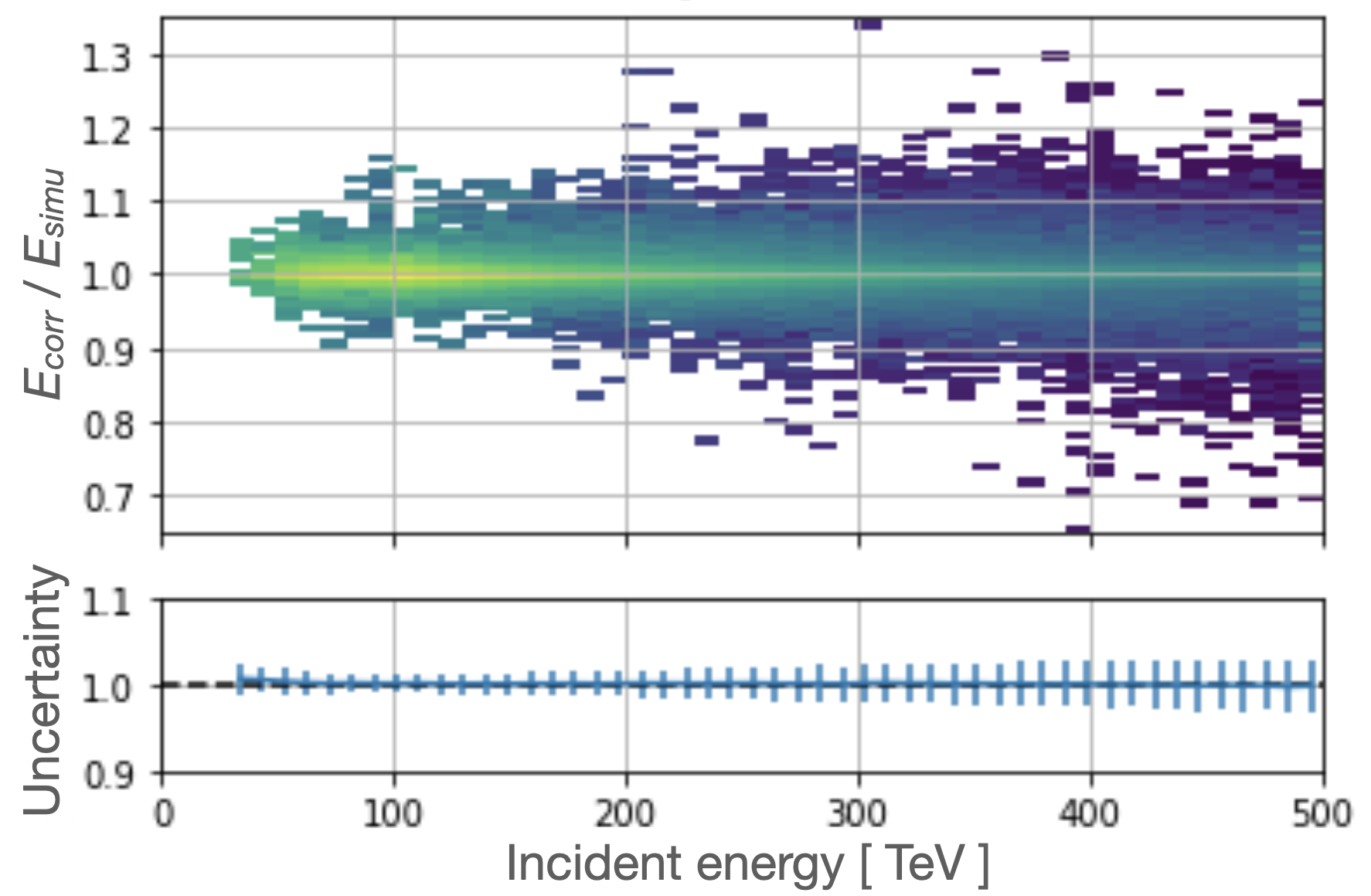}
         \caption{Upper plot: reconstructed BGO deposit energy over its true simulated value in dependence on the incident energy. Lower plot: uncertainty of the BGO deposit energy reconstruction.}
         \label{fig:ecorr_over_esimu}
    \end{minipage}
\end{figure}

The distribution of the total BGO energy before and after saturation correction as well as the true distribution are shown in figure \ref{fig:BGO_energy_distribution}. The no-correction distribution shows a sharp cut-off at BGO energy slightly above 100 TeV. With saturation correction, we are able to recover the right tail of the distribution, which helps to significantly reduce uncertainty on the unfolding of the primary CR spectrum at highest energies \cite{unfolding}. The uncertainty of the reconstructed total deposited energy in the BGO is shown in figure \ref{fig:ecorr_over_esimu}.

The developed model has equal performance for vertical and inclined events, showing no significant dependence on the direction of the incident particle.

\subsubsection{Model performance for the flight data}

To test the developed saturation correction on the flight data we model the saturation for non-saturated events. For this we select helium candidate events without saturation in BGO. We increase the energy deposited in each BGO bar by a factor of $1.2$. Then, the last-layer bars with energy deposit above $2.5$ TeV, we replace with zero energy. For the middle-layer bars we apply such replacement starting from $7$ TeV. The saturation correction is then applied using the two developed models, to these artificially saturated events. In total we have 192 events from which 105 have last layer saturation and 104 have middle layer saturation (we remind that one event can possibly have saturated bars of both classes). The distributions of the difference between the measured deposited energy in a bar and the reconstructed one is shown in figure \ref{fig:target-prediction_data}. We see a nice correspondence between the actual and the reconstructed bar energy for these artificially saturated events.

We then conduct another test: we retrain the models using the helium MC sample while artificially saturating the last-layer BGO bars with energy depositions larger than 2 TeV and the middle-layer bars with energy depositions larger than 6 TeV. These newly trained models are then applied to the non-saturated flight data (the energy shift by 1.2 is not used) that we artificially saturate at the same thresholds: 2 and 6 TeV for the last and middle layers correspondingly. We find the relative deviation of the prediction to be:

\begin{equation}
    \frac{\text{target} - \text{prediction}}{\text{prediction}} = 0.03 \pm 0.17,
\end{equation}

which confirms the conclusion that the saturation correction model gives an unbiased result for the flight data.

\subsubsection{Saturation model for proton, carbon, oxygen and iron Monte-Carlo samples}

\begin{figure}
    \centering
    \begin{minipage}[t]{0.45\textwidth}
         \includegraphics[width=\textwidth]{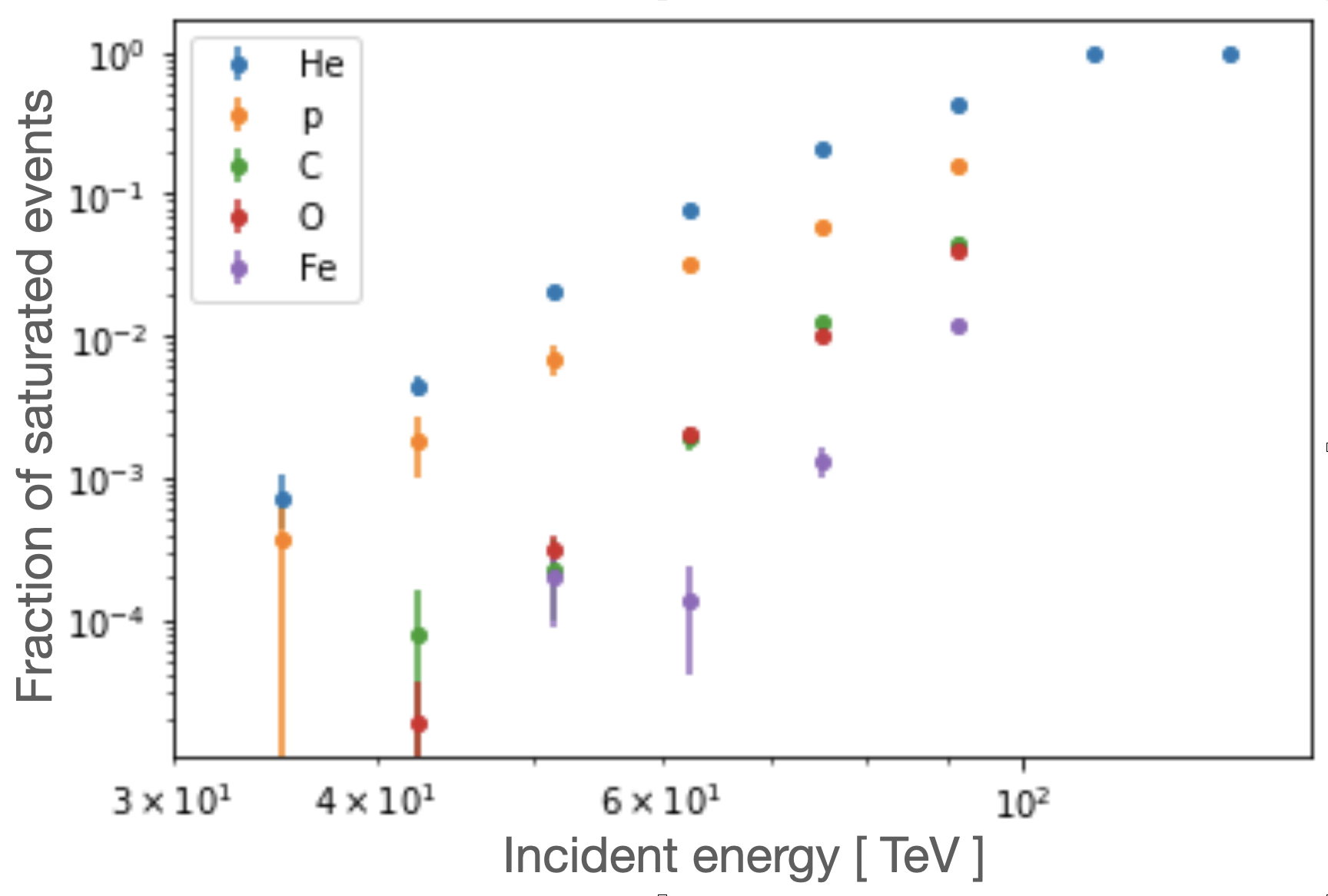}
         \caption{Fraction of saturated events as function of the particle incident energy for proton, helium, carbon, oxygen and iron.}
         \label{fig:frac_sat_phcof}
    \end{minipage}
    \hfill
    \begin{minipage}[t]{0.45\textwidth}
         \includegraphics[width=\textwidth]{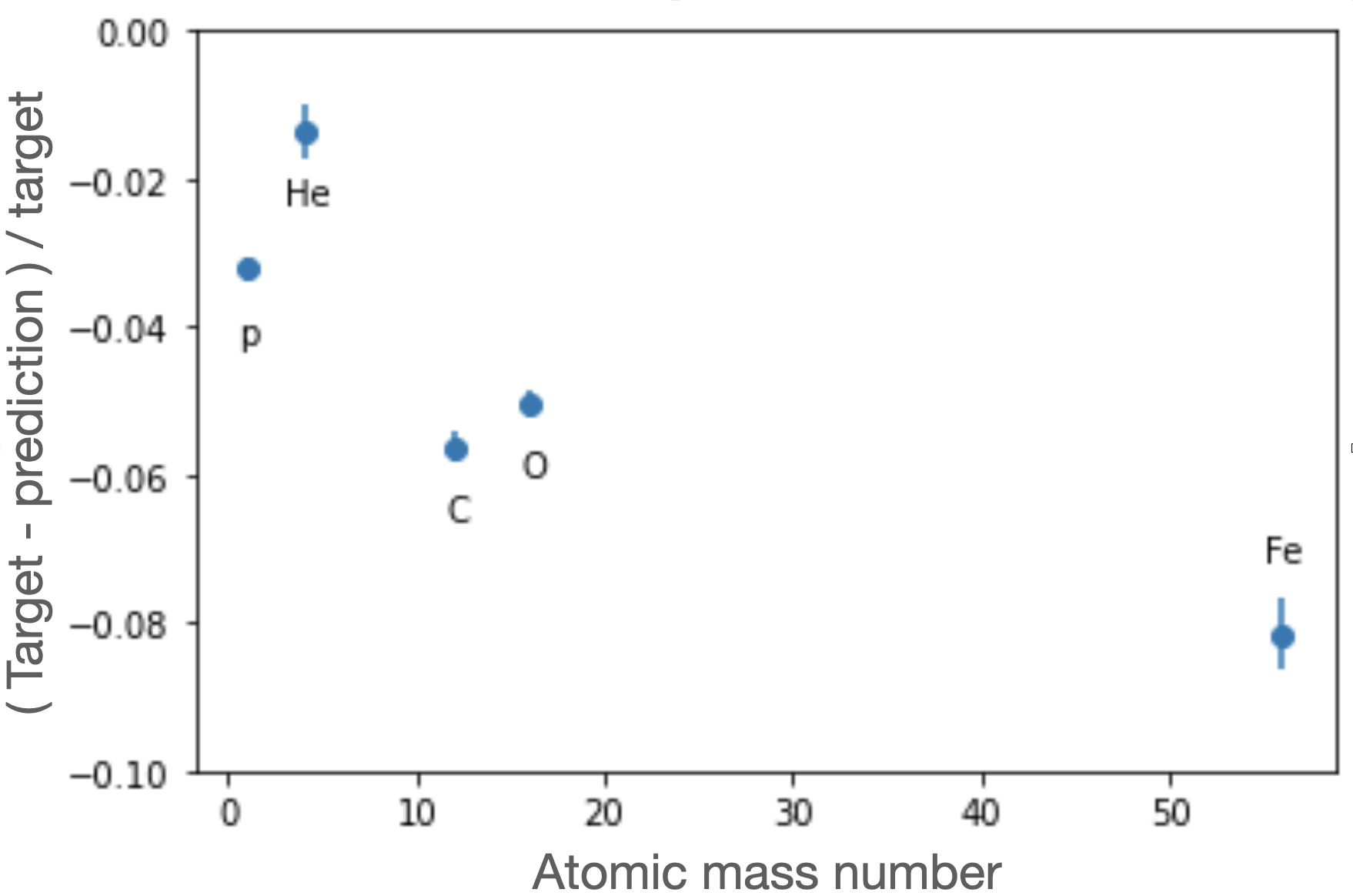}
         \caption{Bias of the helium model prediction as function of the atomic mass number of the incident particle. The vertical error bars correspond to the error of the bias value.}
         \label{fig:a-dependence}
    \end{minipage}
\end{figure}

First, we test the performance of the helium models on the proton, carbon, oxygen and iron MC samples. From the Heitler's model of shower development \cite{heitler}, one can expect that the shower density, and hence the fraction of the saturated events, is lower for the incident particles with large atomic mass number (see figure \ref{fig:frac_sat_phcof}). Interestingly, the fraction of the saturated events for protons is found to be 2-3 times lower than that for helium. This is likely due to the later start of the shower development for protons, such that the shower is not mature enough to saturate at the last BGO layer as much as helium shower. Since the neural network model relies on the general shape of the shower, and since it is different for different ions, the prediction of the models trained on the helium MC sample is biased for other ions, as shown on the figure \ref{fig:a-dependence}.

Since the saturation is much rarer for heavy ions, it is fairly difficult to train a dedicated model for each ion. The statistics of the MC sample used for training would have to be orders of magnitude larger than it is for helium. Instead, we suggest using the helium model with bias correction, as shown in figure \ref{fig:a-dependence}. For protons, it is feasible to train the dedicated model, since the BGO bars saturate for the proton events almost as often as for helium. We developed and trained such a model and made sure that its performance is similar to the model trained for the helium events.


\section{Conclusions and discussion}

An analytical method of computing the energy missing due to saturation of the DAMPE BGO calorimeter was published previously \cite{saturation}. The method reconstructs the energy missing in a BGO bar using the energies deposited in surrounding bars. Thus, by construction, this method can not be applied to events where the saturated bars are adjacent to each other. The method presented in this paper does not have this downside and can be applied to events with any number of adjacent saturated bars. One can see from the figure \ref{fig:frac_sat_adj_iso} that starting from 100 TeV incident energy the events with adjacent saturated bars constitute a significant fraction of the events. Thus the application of the CNN model is not only providing more precise results but also has larger acceptance at highest energies. Quantitatively, for the last layer saturated bars (most frequent saturation case), the current method gives a prediction with a standard deviation about twice smaller than the analytical model across the entire the range of the incident energies.

The developed CNN model shows good performance both on MC and flight data samples. While in the current work we were focused on the saturation for the helium events, we have shown that the similar model is feasible for protons. For the heavier nuclei we suggest to use the helium model and introduce an additional bias correction.

\vspace{6pt}

\acknowledgments{The DAMPE mission was funded by the strategic priority science and technology projects in space science of the Chinese Academy of Sciences. In China, the data analysis was supported in part by the National Key Research and Development Program of China (no. 2016YFA0400200), the National Natural Science Foundation of China (nos. 11525313, 11622327, 11722328, U1738205, U1738207, and U1738208), the strategic priority science and technology projects of the Chinese Academy of Sciences (no. XDA15051100), the 100 Talents Program of Chinese Academy of Sciences, and the Young Elite Scientists Sponsorship Program. In Europe, the activities and the data analysis were supported by the Swiss National Science Foundation (SNSF), Switzerland, National Institute for Nuclear Physics (INFN), Italy and European Research Council (ERC) under the European Union’s Horizon 2020 research and innovation programme (grant agreement No 851103).

The computations presented in this document were performed at University of Geneva on the Baobab cluster, with significant help from computer engineer Y. Meunier and from the HPC team. Simulations were performed on INFN CNAF and ReCaS clusters, Italy, and on Swiss National Supercomputing Centre (CSCS) Piz Daint (project s979).}





\end{document}